\documentclass{iopart}

\usepackage{ifthen}
\usepackage{ifpdf}
\usepackage{color}

\usepackage{latexsym}
\usepackage{amssymb}
\usepackage{bm}

\ifpdf
\usepackage{graphicx}
\usepackage{epstopdf}
\else
\usepackage{graphicx}
\usepackage{epsfig}
\fi
\graphicspath{{./Figs/}{./}}

\newcommand{\rmrk}[1]{#1} 
\newcommand{\Eq}[1]{\textcolor{blue}{Eq.~(\ref{#1})}} 
\newcommand{\Fig}[1]{\textcolor{blue}{Fig.}~\ref{#1}}

\newcommand{\im}{\mbox{Im}}

\newcommand{\eexp}{\mbox{e}^}

\newcommand{\ket}{\rangle}

\newcommand{\tbox}[1]{\mbox{\tiny #1}}
 
\newcommand{\amatrix}[1]{\matrix{#1}}

\newcommand{\mylabel}[1]{\label{#1}} 
\newcommand{\beq}{\begin{eqnarray}}
\newcommand{\eeq}{\end{eqnarray}} 
\newcommand{\be}[1]{\begin{eqnarray}\ifthenelse{#1=-1}{\nonumber}{\ifthenelse{#1=0}{}{\mylabel{e#1}}}}
\newcommand{\ee}{\end{eqnarray}} 

\newcommand{\hide}[1]{}

\begin{document}

\title[Quantum stirring]{Multiple path adiabatic crossing in a 3 site ring}

\author{Dotan Davidovich, Doron Cohen}

\address{
Department of Physics, Ben-Gurion University of the Negev, \\ P.O.B. 653, Beer-Sheva 84105, Israel
}

\begin{abstract}
We find an exact expression for the current 
that is induced in a 3~site ring 
during a multiple-path adiabatic crossing.
The understanding of the dynamics requires 
to go beyond the two-level phenomenology.
In particular we highlight a prototype 
process, ``adiabatic metamorphosis", 
during which current is flowing through 
a non-accessible site. This helps to 
understand the crossover from coherent 
non-classical {\em splitting} to stochastic 
noisy-alike {\em partitioning} of the current.
\end{abstract}

\section{Introduction}

Adiabatic quantum transport \cite{Th1,Th2,Berry,Avron,Robbins} 
is a major theme in quantum mechanics, with diverse applications, 
e.g. quantum Hall effect \cite{hall}, 
dynamics of Josephson junctions \cite{JJ}, 
and the analysis of pericyclic reactions \cite{manz}.  
If a parameter is slowly varied in a closed system 
that has a non-trivial topology, say a ring shaped device,
the formalism implies that current is induced. 
In the absence of magnetic fields we call 
such an effect ``quantum stirring" \cite{pmc,pmt,pml,pms}.
On the one hand it is related to the classical problem 
of ``stochastic stirring" \cite{Saar,st1,st2,st3}, 
and on the other hand it is related to ``quantum pumping" 
in open systems \cite{BPT,pmp1,pmp2,pmp3,pmp4,pmo,pmp5,pmp6}.    

Most results regarding adiabatic quantum transport 
are rather abstract, based on a formal mathematical approach,   
notably the ``Dirac monopoles picture" \cite{Avron,pmc}.
This should be contrasted with the analysis of 
stochastic stirring where the phenomenology is quite mature \cite{Saar}. 
The way to gain better physical insight is to analyze prototype 
model systems \cite{pml,pms}, and to identify the elementary 
ingredients that determine the nature of the dynamics.

\begin{figure}[t!]
\centering
\includegraphics[width=4cm]{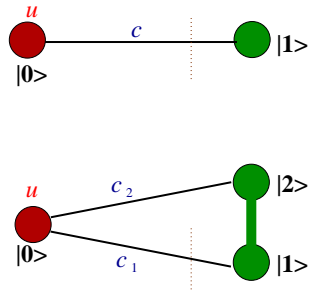}
\caption {
Two-site and three-site toy models for transport. 
A particle is initially positioned at the left site $|0\rangle$, 
called ``dot". The dot has a potential energy~$u$ 
that can be controlled externally. 
As~$u$ is varied adiabatically from $-\infty$ to $+\infty$, 
currents are induced in the bonds, 
and the particle ends up at the right sites.
In the case of a 3-site system it is a multiple-path 
transition through the ${0\leadsto 1}$ and ${0\leadsto 2}$ 
bonds to the lower level of the ``wire".}
\label{f1}
\end{figure}

In the present work we would like to address 
the minimal model for a closed isolated quantum system 
that has a non-trivial topology. This is 
evidently the 3~site ring that is illustrated in \Fig{f1}. 
Quite generally, in the absence of magnetic field, 
the stationary states of the system, and the 
ground state in particular, carry zero current.
If we want to get current we have to drive the 
system by varying a parameter~$u$ in time. 
In the adiabatic limit the current is given by 
the following formula 
\be{1}
\langle \mathcal{I} \rangle  \ \  =  \ \ G(u(t)) \ \dot{u}
\ee
where $G(u)$ is the geometric conductance:
\be{2}
G(u) \ \  = \ \ 
2\im \left[ \Big\langle 
\frac{\partial}{\partial \phi} \Psi \Big| \frac{\partial}{\partial u} \Psi 
\Big\rangle \right]_{\phi=0}
%
%
\eeq
In the above formula $\Psi$ is the adiabatic 
ground state that depends on the parameter~$u$, 
and on an auxiliary test flux $\phi$ through 
the bond of interest.

Specifically we want to consider the following 
scenario, which we call ``Multiple path adiabatic crossing".
Assume that a particle is placed in the 0th site, 
which we call ``dot". The potential of the dot 
is raised slowly from $u=-\infty$ to $u=+\infty$. 
As a result the particle is adiabatically transferred 
from the dot to the other two sites. 
These two sites ("1" and "2") can be regarded 
as a two-orbital entity that we call ``wire".   
At the end of the process the particle will be 
found in the lower energy level of the wire.
We ask what is the current through the first bond ($0 \leadsto 1$).
Equivalently we can characterize the transport 
by the integrated current
\be{3}
Q(u) \ \ = \ \ \int_{-\infty}^u G(u') \ du' 
\eeq     
In particular we define $Q \equiv Q(\infty)$.
If we had single-path geometry obviously
the result for the latter would be ${Q=1}$, 
reflecting 100\% transition probability.
But we are dealing here with a multiple path geometry.

At this point it is important to emphasize 
that if we were dealing with a stochastic 
process the current would be {\em partitioned} 
between the paths, hence ${|Q|<1}$. 
But the essence of ``quantum stirring" is the 
observation that during the driving process 
a circulating current is induced.
Due to this circulation, the integrated current 
can be {\em any} number.

The above recipe \Eq{e2} for calculating adiabatic currents 
is well known from the works on adiabatic transport, 
but its physical implications have not been fully recognized. 
In fact the original interest in this model has been motivated 
by a wrong assertion that ``adiabatic pumping" in a closed system has to be 
quantized \cite{aleiner}. The fallacy of this statement has been 
illuminated using the ``Dirac monopoles picture" \cite{pmc}
and later using a 2~level ``splitting ratio" phenomenology \cite{pms,cnt}. 
The exact solution of the 3~site ring has been 
considered as well \cite{pmc}, to establish that $Q$ of a closed 
driving cycle can have any value.      
However, the full solution of the multiple-path 
adiabatic crossing has not been explored.
In particular it has remained vague whether 
to go beyond the two level approximation is 
of any physical significance.

{\bf Outline.-- } We first derive an exact expression 
for $Q(u)$. This is quite straightforward, but as usual 
the exact expression is not very illuminating physically. 
We therefore try to derive approximations that are based 
on a two-level phenomenology.
Then we realize that there are different regimes 
depending on the ratio between the inter dot-wire and 
the intra-wire couplings. In particular we find a regime 
where the crossing process involves an ``adiabatic metamorphosis" 
stage, during which current is flowing through 
the energetically non-accessible dot.

\section{$G(u)$ for a 2-site system}

We start with the analysis of a single-path crossing in a 2~site system.
The Hamiltonian and the associated current operator are 
\be{4}
\mathcal{H} \mapsto  \left( \amatrix{  u(t) & C^*  \cr  C & u_c } \right), 
\ \ \ \ \ \ \ \
\mathcal{I} \mapsto \lambda\left( \amatrix{ 0 &iC^* \cr -iC &0} \right) \hide{amatrix}
\ee
where $u(t)$ is the potential of the dot, 
and $u_c$ is the level that is crossed,  
and $C$ is the dot-level coupling.
The extra parameter $\lambda=1$ is reserved for later.  
Without loss of generality we assume $C$ to be 
real and positive ${C>0}$.
For the purpose of defining the current 
operator ${\mathcal{I} \equiv -\partial\mathcal{H}/\partial \phi}$, 
and later using \Eq{e2},
one should substitute $C \mapsto C \eexp{i\phi}$. 

For a given value of $u$ the energy of the adiabatic ground state is 
\beq
E(u) \ \ = \ \ \frac{1}{2}\left[(u+u_c) - \sqrt{4C^2 + (u-u_c)^2}\right]
\eeq
The corresponding eigenstate is 
\beq
|\Psi\rangle \ \ \mapsto  \ \ \frac{1}{\sqrt{S}}
\left( \amatrix{  E-u_c  \cr  C\eexp{i\phi}} \right)
\eeq 
where the normalization factor for zero flux is 
\beq
S \ \ = \ \ (E-u_c)^2 + C^2
\eeq
Using \Eq{e2} we get 
\beq
G(u) \ \  =  \ \ C^2 \, \frac{\partial}{\partial u}\left[\frac{1}{S}\right]
\eeq
leading to 
\be{9}
G(u) \ \  =  \ \ \lambda\frac{2C^2}{\left(4C^2 + (u-u_c)^2\right)^{3/2}}
\eeq
where $\lambda=1$. \rmrk{It is easily verified that 
upon integration $Q=\lambda$, hence ${Q=1}$, 
as implied by the continuity equation for 
a single-path adiabatic crossing.}

\section{$G(u)$ for a 3-site system}

We now use the same procedure for the analysis of 
the double-path crossing in a 3~site system.
The Hamiltonian and the associated current operator are 
\be{10}
\mathcal{H}  \mapsto  \left( \amatrix{  u(t) & c_1^* & c_2^* \cr  c_1 & 0 & c_0^* \cr c_2 & c_0 & 0 } \right), 
\ \ \ \ \ \ \ \
\mathcal{I} \mapsto  \left( \amatrix{ 0 &ic_1^* & 0\cr -ic_1 &0 &0 \cr 0 &0 &0 } \right) \hide{amatrix}
\eeq 
We assume the $c$s to be real (no magnetic field) but 
for the purpose of defining the current through the ${0 \leadsto 1}$ bond,  
and later using \Eq{e2}, we substitute $c_1 \mapsto c_1 \eexp{i\phi}$. 

The secular equation for the eigenvalues is 
\beq
E_n^3-uE_n^2-(c_0^2+c_1^2+c_2^2)E_n + c_0^2u - 2c_0c_1c_2\cos(\phi) \ = \ 0
\ee
with the solution
\beq
E_n \ \ = \ \ \frac{u}{3} + 2\sqrt{Q}\cos\left(\frac{\theta}{3} + n\frac{2\pi}{3}\right), \ \ \ \ \ \ n = 0,\pm 1
\ee
where 
\beq
\cos(\theta)&\equiv &\frac{\mathcal{R}}{\sqrt{\mathcal{Q}^3}} \\
\ \ \ \ \ \mathcal{Q} &\equiv & \frac{1}{9}u^2+\frac{1}{3}(c_0^2+c_1^2+c_2^2) \\
\ \ \ \ \ \mathcal{R} &\equiv &\frac{1}{27}u^3+\frac{1}{6}(c_1^2+c_2^2-2c_0^2)u +c_0c_1c_2\cos(\phi) 
\ee
As $u$ is varied from $-\infty$ to $+\infty$, 
the angle $\theta$ varies from $\pi$ to $0$, 
and the ground state energy $E_1$ goes from $-u$ to $-c_0$.
The corresponding eigenstates are   
\be{16}
| n(u) \ket  \ \ \mapsto \ \ 
\frac{1}{\sqrt{S_n}}
\left(\amatrix{E_n^2 - |c_0|^2  & \cr c_1E_n + c_0^*c_2 & \cr c_2 E_n + c_0c_1 } \right)
\ee
The normalization factor for zero flux is 
\beq
S_n &=& (E_n^2 - c_0^2)^2 + (c_1E_n + c_0c_2)^2  + (c_2E_n + c_0c_1)^2 \\
&=& E_n^4 + (c_1^2+c_2^2-2c_0^2)E_n^2 + 2c_0c_1c_2 E_n + c_0^2 (c_0^2 + c_1^2 + c_2^2)
\ee
If we placed the test flux at the $c_0$ bond we would get from \Eq{e2}  
the result that had been derived in \cite{pmc} 
for the current in the $1\leadsto2$ bond, namely:
\beq
G_{1\leadsto2}(u) \ \  =  \ \ 
c_0^2 \, (c_1^2-c_2^2) \, 
\frac{\partial}{\partial u} \left[\frac{1}{S_1}\right]
\eeq
But our interest is in the current that 
goes through the $0\leadsto1$ bond.
Accordingly we have placed the test flux at $c_1$ and get
\beq
\hspace*{-1cm}
G \ = \ 2\left[c_1^2E_1+c_0c_1c_2\right] \frac{1}{S_1}\frac{\partial E_1}{\partial u} 
-\left[c_1^2 E_1^2 + 2c_0c_1c_2E_1 + c_0^2 c_1^2 \right] \frac{1}{S_1^2}\frac{\partial S_1}{\partial u}
\\ \label{e21}
= \frac{d}{du}\left[
\frac{c_1^2 E_1^2 + 2c_0c_1c_2E_1 + c_0^2 c_1^2 }
{E_1^4 + (c_1^2+c_2^2-2c_0^2)E_1^2 + 2c_0c_1c_2 E_1 + c_0^2 (c_0^2 + c_1^2 + c_2^2)}
\right]
\ee

\section{The integrated current}

\begin{figure}
\centering
\includegraphics[clip, width=65mm]{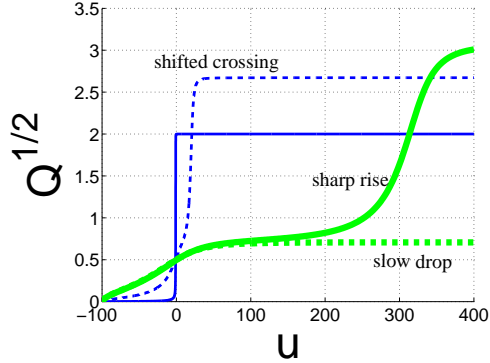}
\caption{
An initially loaded level crosses the other two levels 
of a 3~site network. We plot the parametric variation 
of the integrated current $Q(u)$ in representative cases. 
For graphical purpose the horizontal axis is $Q^{1/2}$.   
The parameters are $c_0=1$ and ${(c_1,c_2)}$ as follows:
Solid blue ${(0.2,0.15)}$ is like a simple 2~level crossing;
Dashed blue ${(5.0,4.3)}$ exhbits a shifted 2~level crossing;
Solid green ${(19,17)}$ features a sharp metamorphosis;
Dashed green ${(19,-17)}$ features a gradual metamorphosis.}
\label{f2}
\end{figure}

\rmrk{On the basis of} \Eq{e21} one observes that 
for any ${c_0\ne 0}$ the integrated current \Eq{e3}
at the end of the process is
\be{22}
\hspace*{-1cm}
Q \ \ = \ \ \left.\frac{c_1^2 E_1^2 + 2c_0c_1c_2E_1 + c_0^2 c_1^2 }{E_1^4 +(c_1^2+c_2^2-2c_0^2)E_1^2 + c_0\cdots}\right|_{E_1=-c_0}
\ \ = \ \ \frac{c_1}{c_1-c_2}
\eeq
Strangely enough this does not depend on the value of $c_0$.
But for ${c_0 = 0}$, \rmrk{based on the {\em same} expression}, 
the result is quite different:
\be{23}
\hspace*{-1cm}
Q \ \ = \ \ \left.\frac{c_1^2}{E_1^2+(c_1^2+c_2^2)}\right|_{E_1=0}
\ \ = \ \ \frac{|c_1|^2}{|c_1|^2+|c_2|^2}
\eeq
It is therefore required to explain what happens {\em physically} 
if $c_0$ is very very small but not zero. In \Fig{f2} 
we illustrate $Q(u)$ for several representative cases. 
If $c_0$ is large $Q(u)$ rises monotonically in a step-like 
fashion to the value that is predicted by \Eq{e22}.
However if $c_0$ is small one observes two stages in the 
parametric evolution: first $Q(u)$ rises to the value
that is predicted to \Eq{e23}, and only after that it re-adjust 
to the value of \Eq{e22}. 
\rmrk{In Section~9 we shall use the term ``adiabatic metamorphosis"
in order to describe this re-adjustment of the occupations. We shall see that it 
involves a much larger parametric scale ${u_m \propto 1/c_0}$ 
that diverges in the limit ${c_0\rightarrow0}$.}
Hence for ${c_0 = 0}$ we are left with \Eq{e23} \rmrk{instead of} \Eq{e22}.  
\rmrk{A closer inspection of the metamorphosis stage} 
(dashed vs solid green curves in \Fig{f2}) reveals 
that it can be either a gradual or a sharp transition, 
depending on the relative sign of~$c_1$ and~$c_2$.  
 
\begin{figure}
\centering
\includegraphics[clip, width=60mm]{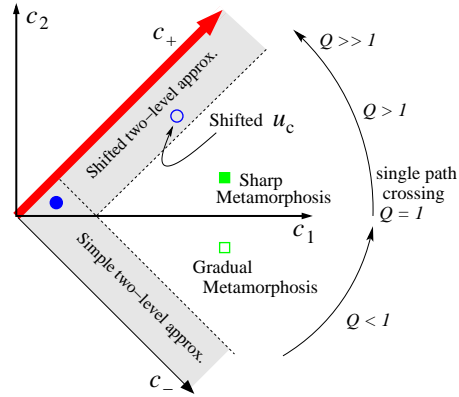}
\caption{
A schematic diagram that shows the different regimes 
in the analysis of the adiabatic double-path crossing.
Setting ${c_0=1}$ the parameters that define 
the 3-site system are the couplings $c_1$ and $c_2$.  
Without loss of generality we relate to one quarter where both $c_{+}$ and $c_{-}$ are positive. 
Grey shading indicate the regime where a two level approximation
scheme can be used, either $c_{-}\ll c_0$ or $c_{+}\ll c_0$, as discussed in the text.  
In each regime the $G(u)$ has a different looking lineshape.  
Blue and green symbols indicate the representative illustrations 
that have been displayed in \Fig{f2}.}
\label{f3}
\end{figure}

The values $(c_1,c_2)$ for the illustrations in \Fig{f2}
are indicated in the diagram of \Fig{f3}. In the following 
sections we would like to illuminate the different regions 
in this diagram by attempting a two-level approximation scheme.

\section{The 2-level approximation}

Let us try to reduce the 3~level dynamics to a 2~level crossing problem. 
For this purpose we switch to the following basis:
\beq
|0\ket \ \ &=& \ \ \mbox{the dot state}
\\ 
|+\ket \ \ &=& \ \ \frac{1}{\sqrt{2}}(|1\ket + |2\ket) \ \ = \ \ \mbox{the upper (even) wire state}
\\
|-\ket \ \ &=& \ \ \frac{1}{\sqrt{2}}(|1\ket - 2\ket) \ \ = \ \ \mbox{the lower (odd) wire state}
\eeq
In the new basis the Hamiltonian and the current operator \Eq{e10} take the following form:
\be{27}
\mathcal{H} \mapsto \left( \amatrix{   \rmrk{u(t)} &c_+ &c_-\cr c_+ &c_0 &0 \cr c_- &0 &-c_0 } \right), 
 \ \ \ \ \ \ \ \
\mathcal{I} \mapsto \frac{c_1}{\sqrt{2}} \left( \amatrix{   0 &i & i\cr -i &0 &0 \cr -i &0 &0 } \right) 
\eeq
with couplings
\beq
c_{\pm} \ \ = \ \ \frac{1}{\sqrt{2}} (c_1 \pm c_2)
\ee
Without loss of generality we focus on the strongest bond, 
meaning that we assume ${|c_1|>|c_2|}$, and by appropriate gauge 
we arrange that ${c_1>0}$, hence both $c_{\pm}$ are positive numbers.
We shall see that a two-level approximation scheme 
is useful for the treatment of 3~cases that are indicted in \Fig{f3}, namely
\beq
|c_{+}| \ \ \ll \ \ c_0  \\
|c_{-}| \ \ \ll \ \ c_0 \\
c_0 \ \ = \ \ 0
\eeq
In all these cases we can fit the exact result \Eq{e21} 
to the two-level expression \Eq{e9},  
with some effective values for $C$ and $u_c$ and $\lambda$. 
The remaining case of having a relatively {\em small but finite} $c_0$ is excluded, 
because it cannot be treated within the framework of a two level approximation.
This last case will be considered separately.

\section{The simple 2-level approximation ${|c_{+}| \ll c_0}$}

We first consider the very simple case, in which the third (upper) level  
can be ignored. The condition for that is ${|c_{+}| \ll c_0}$. 
Taking the relevant block from the $3\times3$ Hamiltonian of \Eq{e27}  
one obtains a reduced  $2\times2$ Hamiltonian that is given by \Eq{e4} 
with the effective parameters
\beq
\lambda = \frac{c_1}{c_1-c_2},
\ \ \ \ \ \ 
C = \frac{c_1 - c_2}{\sqrt{2}} ,
\ \ \ \ \ \ 
u_c = -c_0
\ee
Hence we deduce that $G(u)$ of \Eq{e9} can be used 
as an approximation for the exact result. 
This expectation is confirmed in \Fig{f4}. 
 
The presence of $\lambda$ reflects that the flow is via two bonds instead 
of via only a single bond, unlike the case of the 2-sites model. 
In other words the particle ``splits" and flows through both bonds.
We notice that the integrated current $Q=\lambda$ can have a manifestly 
non-classical value: it can be either larger than~1 or negative.
In fact, if it is larger than~1 in one bond it has to be 
negative in the other bond, since the total corresponds 
to 100\% probability of being adiabatically transferred.  
The non-classical value of $\lambda$ reflects a circulating current 
that is induced during the transition. In a semi-classical language 
it means that the particle is looping several rounds 
through the ring before ending up in the wire.

\section{The shifted 2-level approximation ${|c_{-}| \ll c_0}$} 

The simple two-level approximation of the previous section 
is not valid if $|c_{+}| > c_0$. Still if ${|c_{-}| \ll c_0}$ 
we can still get a result that looks like \Eq{e9}.
The solution procedure involves two steps. In the first step
we switch to a new basis:
\be{0}
|\theta\ket \ \ &=& \ \ \cos(\theta/2) |0\ket + \sin(\theta/2) |+\ket
\\
|\bar{\theta}\ket \ \ &=& \ \  -\sin(\theta/2) |0\ket + \cos(\theta/2) |+\ket
\\
|-\ket  \ \ &=& \ \ \mbox{the lower (odd) wire state} 
\ee
where
\be{36}
\theta(u) \ \  =  \ \ \arctan\left(\frac{2c_+}{u-c_0}\right)
\ee
In this basis the block of the Hamiltonian \Eq{e27} that contains 
the strongly interacting states $|0\ket$ and $|+\ket$  
becomes diagonal. Now it is possible to neglect the upper level $|\theta\ket$ 
and we get a two-level crossing problem that involves 
the ``dressed" dot level $|\bar{\theta}\ket$ and the lower (odd) wire level $|-\ket$.  
The adiabatic energy of the former is 
\beq
E_{\bar{\theta}} \ \ = \ \ \frac{1}{2}\left[(u+c_0) - \sqrt{4c_{+}^2 + (u-c_0)^2}\right]
\eeq
while that of the latter is $E_{-} = -c_0$. 
Accordingly the shifted crossing point is 
\be{38}
u_c \ \ = \ \ \left[-1 + \frac{1}{2}\left(\frac{c_{+}}{c_0}\right)^2\right] \, c_0
\ee
The stronger the coupling $c_{+}$ to the upper level, 
the larger is the shift of~$u_c$.

\begin{figure}

(a1) \includegraphics[clip,width=4cm]{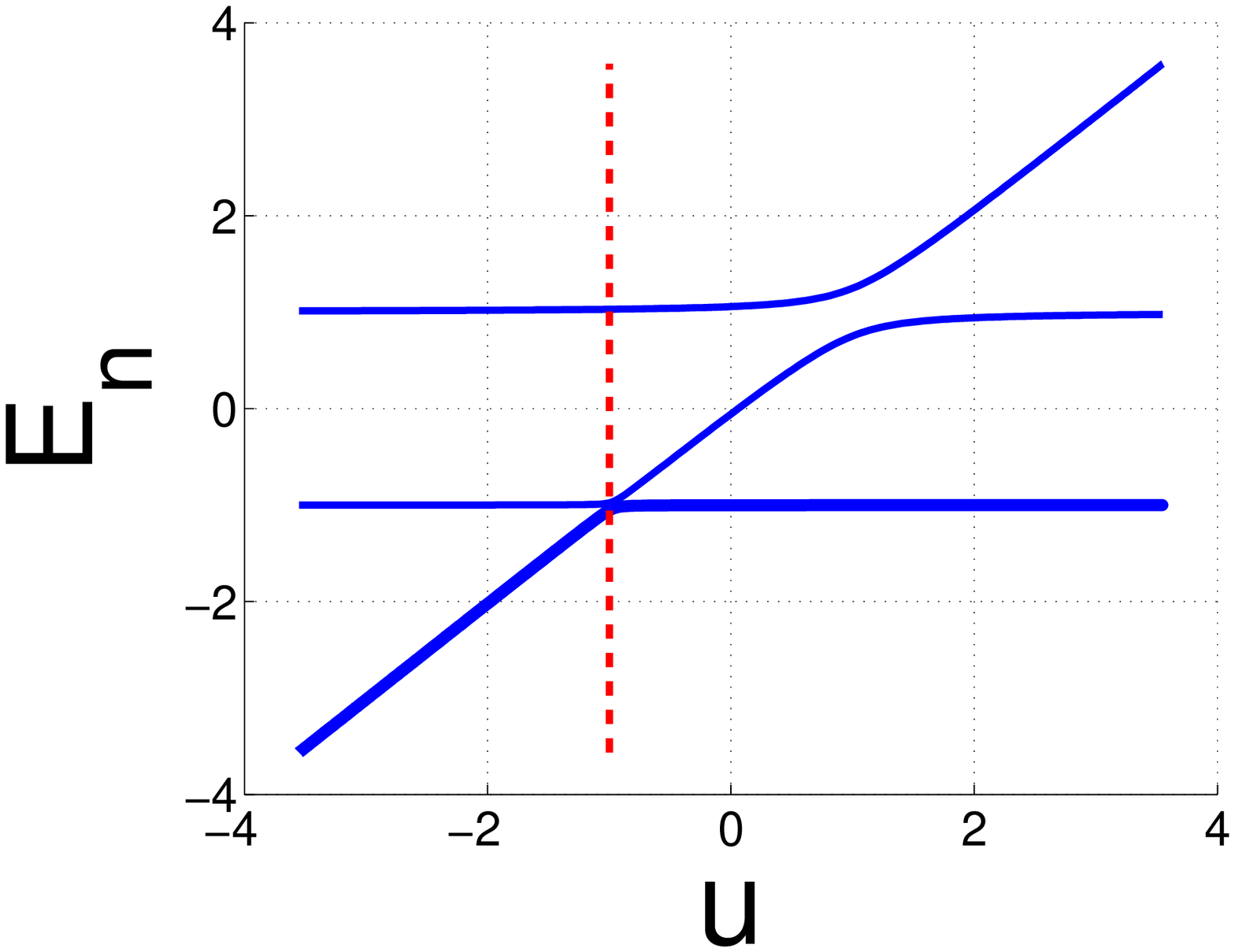}
\ \ \ \ \ \ 
(a2) \includegraphics[clip,width=4cm]{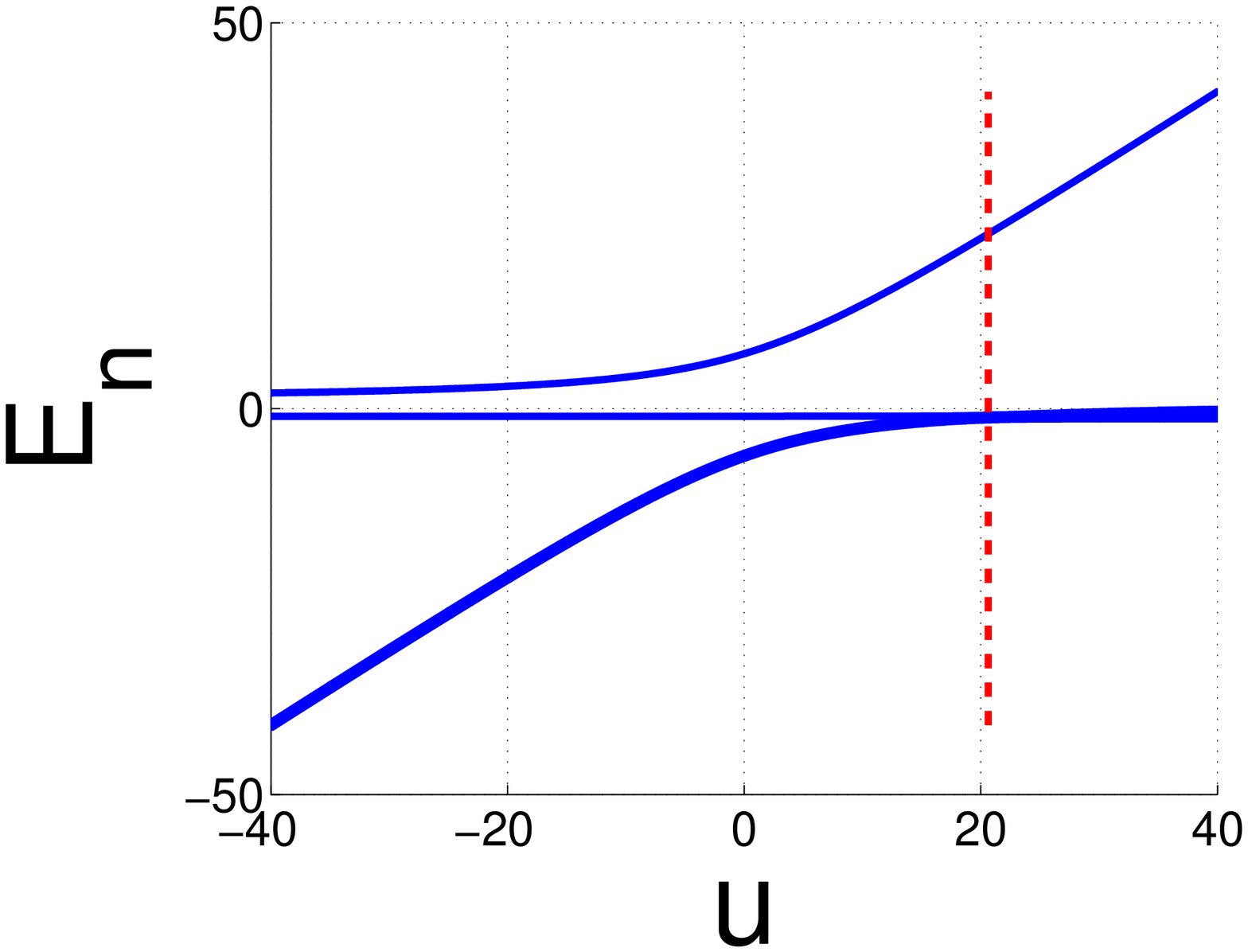}

(b1) \includegraphics[clip,width=4cm]{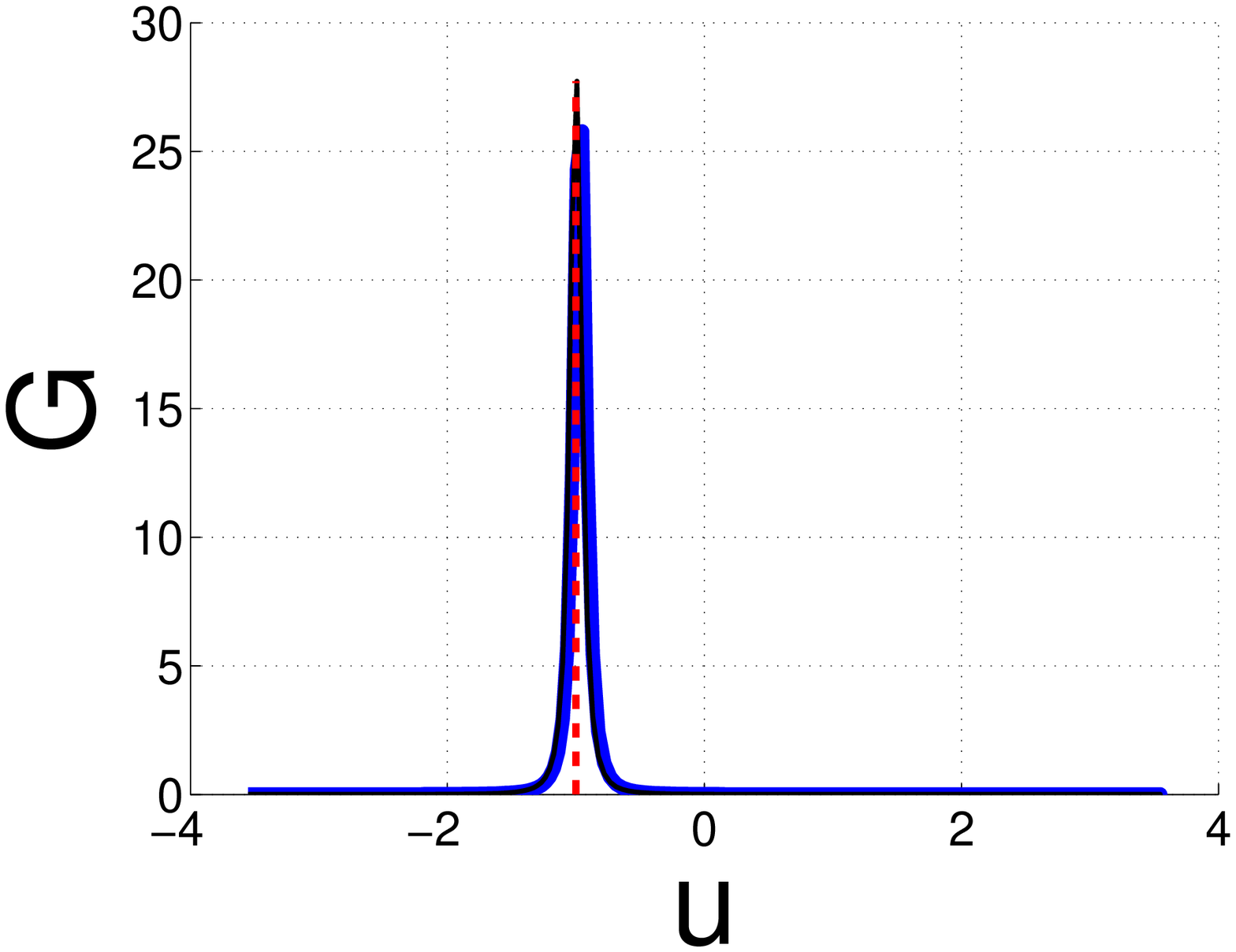}
\ \ \ \ \ \ 
(b2) \includegraphics[clip,width=4cm]{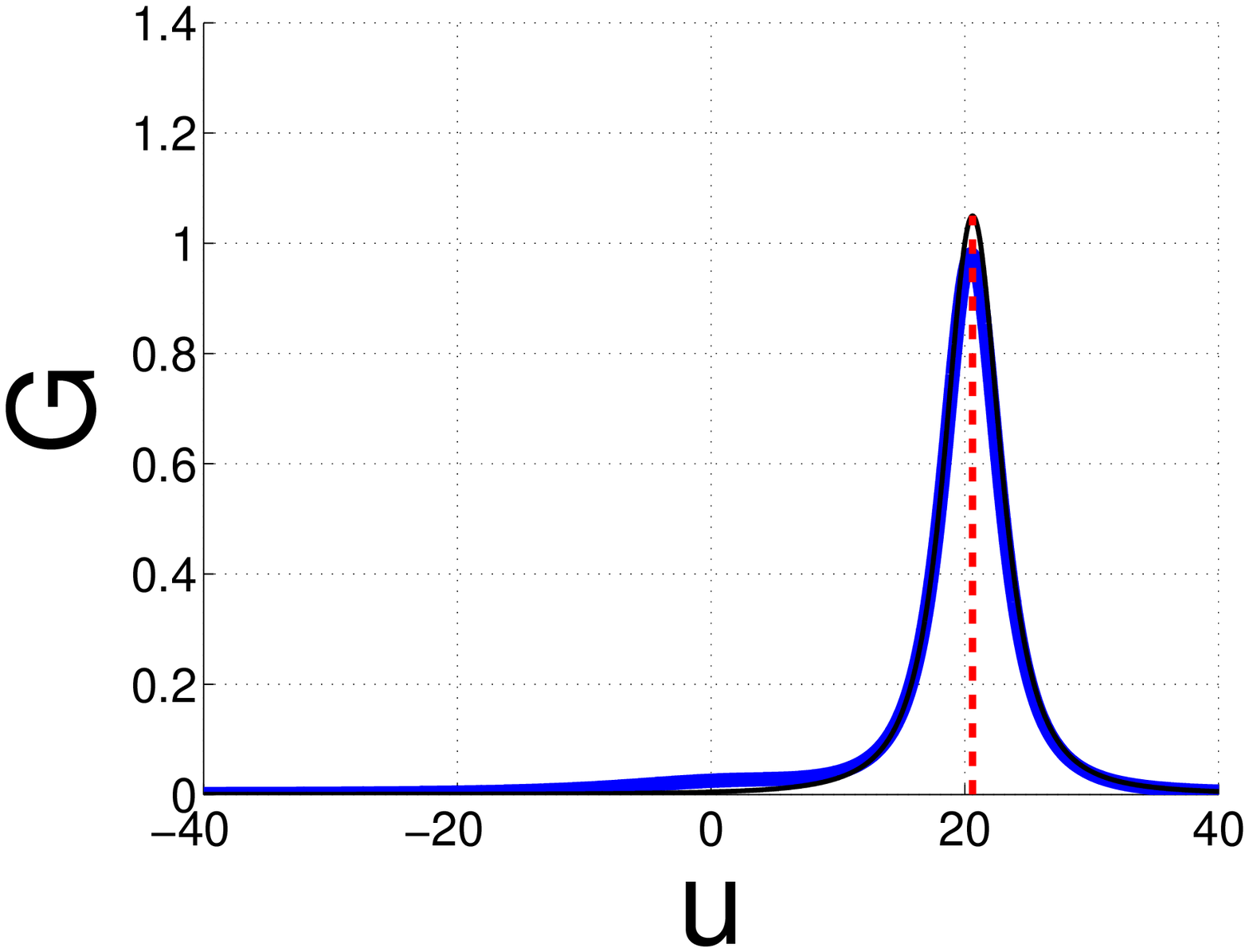}

\caption{
An initially loaded level crosses the other two levels 
of a 3~site network. (a)~The adiabatic energies $E_n(u)$
as a function of the dot potential.
(b)~The geometric conductance $G(u)$ during this sweep 
process, reflecting the current through the $c_1$ bond. 
The thick blue line is the exact solution \Eq{e21}.  
We use units such that ${c_0 = 1}$.
The parameters in set~(1) 
are $c_1 = 0.2$ and $c_2 = 0.15$, 
corresponding to the regime ${|c_{+}| \ll c_0}$, 
where the simple two level approximation (thin black line) applies.
Vertical dashed line indicates the dot-wire crossing point. 
The parameters in set~(2) 
are $c_1 = 5.0$ and $c_2 = 4.3$,
corresponding to the regime $|c_{-}| \ll c_0$, 
where a shifted two level approximation (thin black line) applies.
Vertical dashed line indicates the {\em shifted} crossing point. 
\rmrk{Note: the agreement is so good that the thin black lines 
almost cannot be resolved.}
}
\label{f4}                                                                                

\end{figure}

In order to estimate the effective parameters of the shifted 
two level crossing we have to write the Hamiltonian and the 
current operator in the new basis. In the vicinity of the 
crossing point we set $\theta_c=\theta(u_c)$, getting
\beq  
\hspace*{-15mm}  
\mathcal{H} \ \ \mapsto \ \ 
\left(\amatrix{  
\left[u\sin^2(\theta_c/2) + c_0\cos^2(\theta_c/2)- c_+\sin(\theta_c)\right] 
& \ \ -c_-\sin(\theta_c/2)  \cr  
-c_-\sin(\theta_c/2)  & -c_0 } 
\right) 
\eeq
and
\beq
\hspace*{-15mm} 
\mathcal{I} \ \ \mapsto \ \ \lambda
\left( 
\amatrix{  0 & -ic_-\sin(\theta_c/2)  \cr  ic_-\sin(\theta_c/2) & 0 } 
\right)        
\eeq
where $\lambda$ is the same as in the previous section.
It is important to realize that up to constant
the effective dot potential equals $\alpha u$ with $\alpha=\sin^2(\theta_c/2)$.
It is not difficult to see that this implies the replacement  
\beq
G(u) \ \ \mapsto \ \ \alpha \ G(\alpha u)
\eeq
Hence within the framework of the two level 
approximation the effective $C$ in \Eq{e9}
is not $-c_{-} \sin(\theta/2)$ 
but rather it should be divided by $\alpha$.
So eventually we deduce that the $G(u)$ 
can be approximated by \Eq{e9} with an 
effective coupling parameter 
\beq
C \ \ = \ \ -\left[ \sin\left(\frac{\theta_c}{2}\right) \right]^{-1} \, c_{-}
\eeq
This expectation is confirmed in \Fig{f4}. 
Note again that $\lambda$ is the same as in the simple 
two-level approximation, and that we evaluate $\theta_c$  
at the crossing point using \Eq{e36} with \Eq{e38}.

\section{Adiabatic crossing for $c_0 = 0$}

In the previous versions of the two-level approximation, 
the intra-wire coupling~$c_0$ was large in some sense. 
Now we go to the other extreme limit of having ${c_0=0}$.
\rmrk{This resembles the standard setup that is used in 
the analysis of stimulated Raman adiabatic passage (STIRAP). 
In fact we can adopt here the same ``dark state" picture 
in order to reduce the problem to a two-level crossing.}
Namely, for this purpose we switch to the following basis:
\be{0}
|0\ket \ \ &=& \ \ \mbox{the dot state}
\\
|C\ket \ \ &=& \ \ \frac{1}{\sqrt{c_1^2+c_2^2}}(c_1|1\ket + c_2|2\ket)
\\ 
|D\ket \ \ &=& \ \  \frac{1}{\sqrt{c_1^2+c_2^2}}(c_2|1\ket - c_1|2\ket) \ \ = \ \ \mbox{dark state}
\ee
In the new basis the $|D\ket$ state decouples, 
and hence we end up again with 
a reduced $2\times2$ Hamiltonian 
that is given by \Eq{e4} 
with the effective parameters
\beq
\lambda = \frac{c_1^2}{c_1^2+c_2^2}, 
\ \ \ \ \ \ 
C = \sqrt{c_1^2+c_2^2},
\ \ \ \ \ \ 
u_c = 0
\ee
Hence we deduce that $G(u)$ of \Eq{e9} with the above set 
of effective parameters coincides in this case with the exact result. 

\rmrk{It should be clear that for $c_0=0$ 
we no longer have a non-trivial geometry, 
and hence a circulating current cannot be induced.
For this reason it is a-priori expected 
to get an effective two-level description with ${\lambda \in [0,1]}$. 
In fact we got for $\lambda$ a stochastic look-alike expression 
that reflect the relative transmission of the two bonds.}

\begin{figure}

(a) \includegraphics[clip,width=5cm]{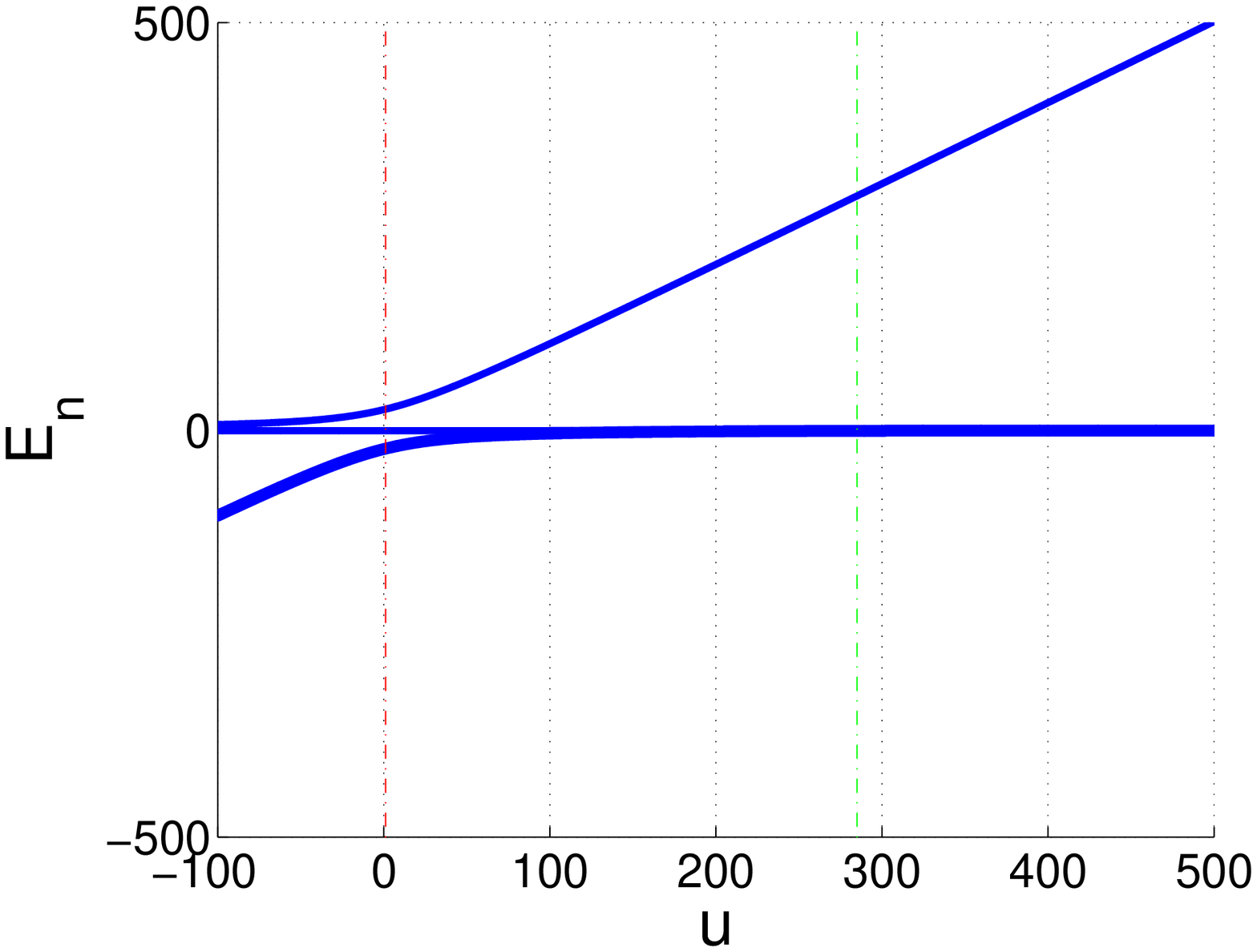}

(b) \includegraphics[clip,width=5cm]{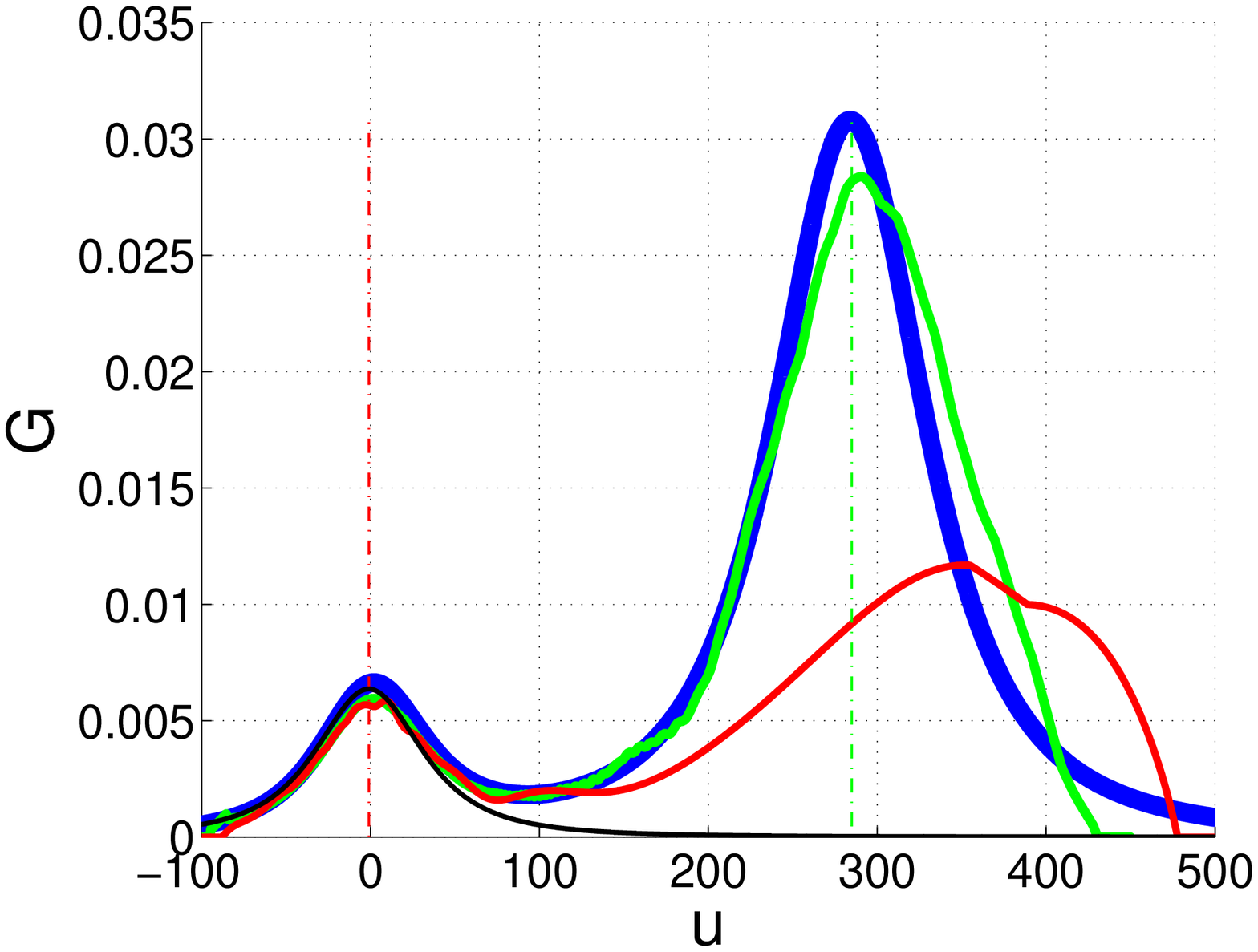} 

(c) \includegraphics[clip,width=5cm]{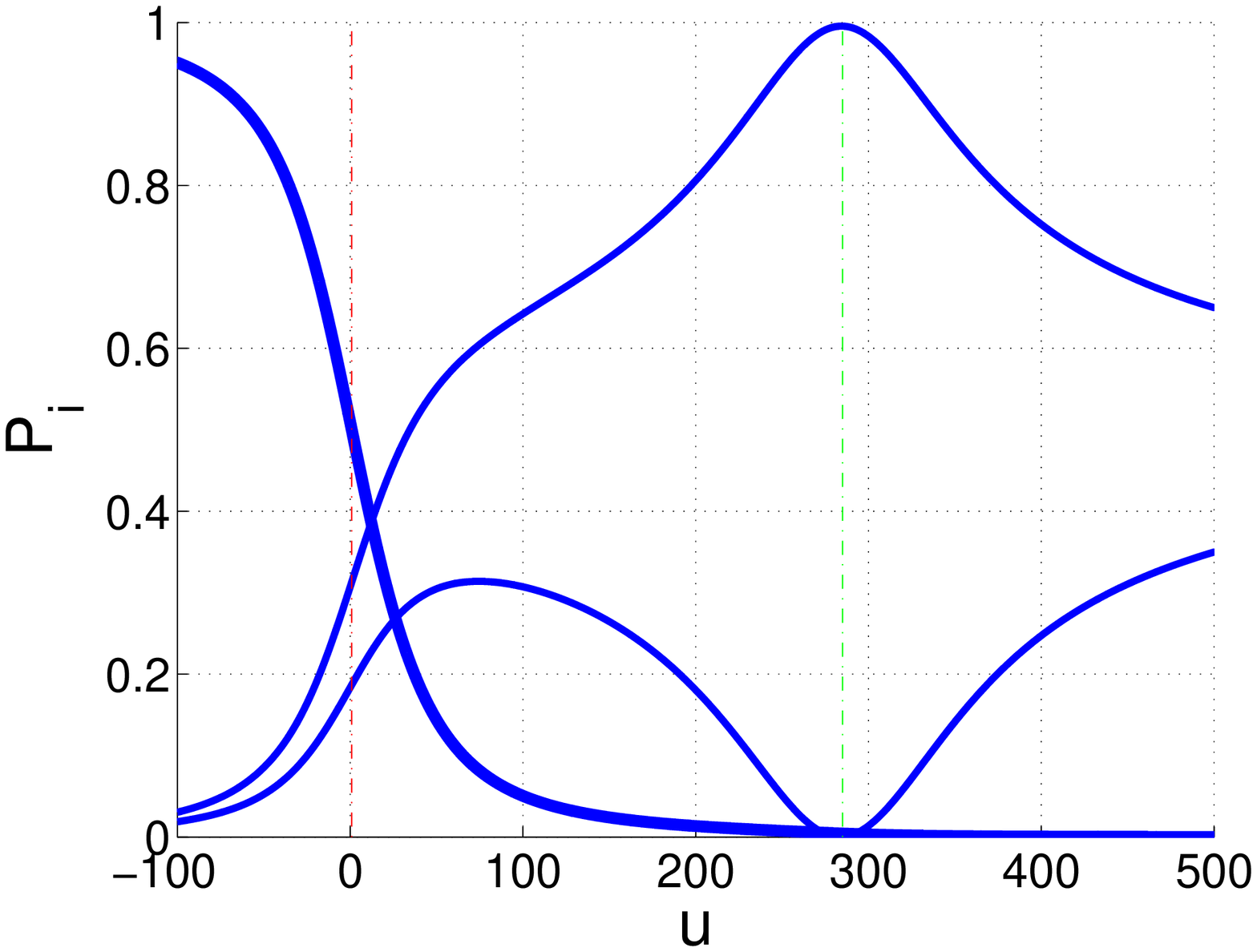}

\caption{
The same as in \Fig{f4} but with $c_1 = 19$ and $c_2 = 15$, 
illustrating a sharp metamorphosis. 
The additional panel (c) shows the parametric 
variation of the occupation probabilities. 
In (b) the black line is the $c_0=0$ solution. 
The thinner (green) and the thinnest (red) lines are $I/\dot{u}$ 
for $\dot{u}=2$ and for $\dot{u}=50$, 
as determined by numerical simulation.
The left and right vertical lines indicate 
the dot-wire crossing point, and the metamorphosis point, 
with separation ${(u_m-u_c)=286}$.  
During the adiabatic metamorphosis a current 
is flowing through the energetically distant dot. 
Mild non-adiabaticity spoils the metamorphosis 
without affecting the dot occupation.
}
\label{f5}
\end{figure}

\section{Adiabatic metamorphosis}
  
Let us contrast the $c_0 = 0$ case with the $c_0 \rightarrow 0$ case. 
The two cases give very different results. We would like to better 
clarify what really happens if $c_0$ is very small.  
We first recall the optional bases for the representation of the system. 
The standard basis is $|0\ket  ,|1\ket , |2\ket$. 
The wire-eigenstates basis is $|0\ket  ,|+\ket , |-\ket$, 
and the $c_0=0$ basis is $|0\ket  ,|C\ket , |D\ket$.    
For the instantaneous eigenstates we shall 
use the notations $|E_g\ket$, and $|E_d\ket$ and $|E_e\ket$. 

Recall that for $c_0=0$ the dark state $|D\ket$  decouples, 
meaning that $|E_d\ket =|D\ket$, while the 
other two eigenstates  $|E_g\ket$ and  $|E_e\ket$
are superpositions of  $|0\ket$ and $|C\ket$. 
At the end of an adiabatic process the system 
will be found in the degenerate $|E_g\ket = |C\ket$ state.
However, if $c_0$ is non-zero, the system 
ends up in the non-degenerate $|E_g\ket = |-\ket$ state.   

It is therefore clear that for very small but finite $c_0$ the 
adiabatic ground-state changes from  $|C\ket$ to  $|-\ket$.
We call this ``adiabatic metamorphosis". 
We define $u_m$ as the value of $u$ at which this 
metamorphosis occurs. Close to $u_m$ the dot level 
is energetically far above, hence the lower 
states $|1\ket$ and $|2\ket$ form 
a 2-level system with virtual coupling through 
the distant dot level. The reduced Hamiltonian 
is determined by 2nd order perturbation theory:
\be{0}
\mathcal{H} \ \ \mapsto \ \  
\left( \amatrix{ 
\frac{c_1^2}{u}  &&    c_0 - \frac{c_1 c_2}{u}   \cr
 c_0 - \frac{c_1 c_2}{u}  &&   \frac{c_2^2}{u}
} \right) \hide{amatrix}
\ee
By inspection of this Hamiltonian it is clear that 
for large enough $u$ the direct coupling $c_0$ takes 
over, and then the metamorphosis to $|E_g\ket = |-\ket$
is finalized. In particular it is interesting 
to consider the case in which ${c_2 \sim c_1}$. 
Then the metamorphosis crossing point is sharply defined:
\beq
u_m \ \ = \ \ \frac{c_1c_2}{c_0},
\ \ \ \ \ \ \mbox{[for sharp metamorphosis]}
\eeq
which is demonstrated in \Fig{f5}.
Otherwise the metamorphosis is a gradual 
process, as was illustrated in \Fig{f2}. 
Note that for sharp metamorphosis, at $u=u_m$, 
all the probability is concentrated in one site 
of the wire, namely, in the site that is more strongly 
connected to the dot. This is demonstrated in panel~(c) of \Fig{f5}.

\section{Beyond the adiabatic limit}

In order for the process to be adiabatic the probability distribution should change slowly 
with time. This implies that the current cannot be very large. Let us see what 
is the precise statement. Specifically for a two level Landau-Zener crossing 
the adiabatic condition is 
\be{49}
\dot{u} \ \ \ll \ \ C^2 
\ee
where $C$ is the coupling between the dot and the crossed level. 
This implies that 
\be{0}
I_{\tbox{max}} \ \ \sim \ \ G(u_c) \ \dot{u}_{\tbox{max}} \ \ \sim \ \ C
\ee
One observes that the maximal current reflects the coupling.
Furthermore, also the integrate current cannot be 
too large. It is simply bounded by unity (${|Q|<1}$)
reflecting that the maximum transfer is 100\%.

If we consider multiple path geometry it is easy 
to show that the conclusion regarding the maximal current 
still holds. However, as was clarified in previous section, 
the integrated current $Q=\lambda$ becomes arbitrarily
large if $c_1 \sim c_2$, rather than being bounded.
A large value ${|Q|>1}$ reflects the existence of 
a circulating current that is induced in the system 
during the driving process. \rmrk{It should be clear 
that in order to witness ${|Q| \gg 1}$ one has to 
satisfy a very demanding adiabatic condition, 
because the effective coupling that enters into \Eq{e49} 
is $C = (c_1{-}c_2)\sqrt{2}$.} 
 
Within the conventional framework of the two-level approximation, 
the implication of non-adiabaticity is to have less than 100\% 
probability to cross from the dot to the wire, 
as implied by the Landau-Zener expression \cite{zener1,zener2}.
But for very small $c_0$, such that the metamorphosis scenario applies, 
the implications of non-adiabaticity are more interesting \rmrk{as explained below}.

\rmrk{For finite $\dot{u}$ there is a finite range of small $c_0$ values 
for which the result for~$Q$ is approximately the 
same as for $c_0 = 0$. The complementary statement is as follows:}
for a given $c_0$, if $\dot{u}$ is large enough, 
the system does not have enough time to realize that it is coupled 
to a ``dark state".  Roughly this non-adiabatic condition 
takes the form $\dot{u}>c_0^2$. 
\rmrk{Thus we have an intermediate ``diabatic" regime ${c_0^2 \ll \dot{u} \ll C^2}$ 
where the dynamics is ``adiabatic" with regard to the crossing, 
but ``sudden" with regard to the metamorphosis.}
The bottom line of the above discussion is illustrated in \Fig{f5}.
One observes that for mild values of $\dot{u}$ the metamorphosis 
stage is not expressed, and the dynamics looks like that of ${c_0=0}$ system.

\section{Discussion}

Transport in quantum networks is a theme that emerges 
in diverse contexts. The simplest network that has non-trivial
topology is the 3~site system that we have considered 
in this paper. It can be regarded as composed of ``dot"
and ``wire" segments. 
The most elementary process that has to be understood is 
an adiabatic sweep of the potential energy of 
a selected site (the dot), leading to transfer 
of the probability to the other sites (the wire). 
Unlike stochastic process in which the probability current 
is partitioned with branching ratios that are bounded within ${[0,1]}$, 
here the splitting ratio $\lambda$ can be any number 
reflecting a quantum stirring effect.  

The detailed analysis of the  3~sites model  
allows to highlight several essential ingredients 
in the analysis of quantum transport. 
In particular it was important to clarify 
what is the way in which the two-level approximation 
breaks down. Strangely enough the splitting 
ratios are independent of the intra-wire coupling, 
but still the $G(u)$ line shape is strongly 
influenced.   

In particular we have distinguished between two type of processes: 
inter dot-wire ``adiabatic crossing" processes;   
and intra-wire ``adiabatic metamorphosis" processes. 
In the former probability is transported between 
the dot and the wire, while in the latter 
the changes in the occupation are exclusively within the wire. 
During the metamorphosis stage the dot level 
is very far from the wire levels, 
but still current flows through the inter-connecting bonds, 
without being accumulated in the dot. 
 
We believe that the processes that we have illuminated 
are of much relevance, and might shed new light,  
on the analysis of pericyclic reactions \cite{manz}.  
In this context the method of calculating 
electronic quantum fluxes had assumed that the latter 
can be deduced from the continuity equation. 
Such procedure is obviously not applicable for 
(say) a ring-shaped molecule: due to the multiple path 
geometry there is no obvious relation between 
currents and time variation of probabilities.
 
Furthermore, it is important to understand how
non-adiabaticity and decoherence affect adiabatic transport.
Possibly the most dramatic demonstration concerns 
the suppression of metamorphosis processes by 
mild non-adiabaticity. Then we get instead of 
coherent splitting, stochastic-like partitioning 
of the current. The reason for this crossover can 
be optionally explained using a very general paradigm.
Namely, once the intra-wire couplings are introduced, 
there is a protecting ``gap" that becomes effective 
if the rate of the sweep is slow enough; 
this protecting gap forces the particle to be in  
a definite superposition at any moment.
It follows that coherent splitting, unlike 
``partitioning" of current is not a noisy process.
This observation has implications on the calculation 
of ``counting statistics" and ``shot noise" \cite{count,cnt}.


\ack
This research has been supported by the Israel Science Foundation (grant No.29/11).


\clearpage
\end{document}